# Atomic-Scale Defect Detection by Nonlinear Light Scattering and Localization

Farbod Shafiei [1,*], Tommaso Orzali [2], Alexey Vert [2,3], Mohammad-Ali Miri [4,5], P. Y. Hung [2], Man Hoi Wong [2], Andrea Alù [4,6,7], Gennadi Bersuker [8], Michael C. Downer [1]

[1] Department of Physics, The University of Texas at Austin, Austin TX USA
[2] SEMATECH, Albany NY USA
[3] Sunny Polytechnic Institute, Albany NY USA
[4] Department of Electrical and Computer Engineering, The University of Texas at Austin, Austin TX USA
[5] Department of Physics, Queens College of the CUNY, Queens NY USA
[6] Photonics Initiative, Advanced Science Research Center, CUNY, New York NY USA
[7] Physics Program Graduate Center, CUNY, New York NY USA
[8] The Aerospace Corporation, Los Angeles CA USA

*Correspondence to: farbod@physics.utexas.edu

**ABSRTRACT:**
**Hetero-epitaxial crystalline films underlie many electronic and optical technologies but are prone to forming defects at their hetero-interfaces. Atomic-scale defects such as threading dislocations that propagate into a film impede the flow of charge carriers and light degrading electrical/optical performance of devices. Diagnosis of sub-surface defects traditionally requires time-consuming invasive techniques such as cross–sectional transmission electron microscopy. Using III-V films grown on Si, we have demonstrated noninvasive, bench-top diagnosis of sub-surface defects by optical second-harmonic scanning probe microscope. We observed a high-contrast pattern of sub-wavelength "hot spots" caused by scattering and localization of fundamental light by defect scattering sites. Size of these observed hotspots are strongly correlated to the density of dislocation defects. Our results not only demonstrate a global and versatile method for diagnosing sub-surface scattering sites but uniquely elucidate optical properties of disordered media. An extension to third harmonics would enable irregularities detection in non-$\chi^{(2)}$ materials making the technique universally applicable.**

**INTRODUCTION:**
III-V epitaxial films are excellent candidates for high-speed electronic and opto-electronic devices, semiconductor spintronics, nanometer scale logic transistors, nanowire channels, and light emitters on the industry-standard Si (001) platform [1-3]. Due to their superior carrier mobility, III-V carrier transport channels can improve the performance of silicon-based logic transistors by enabling reduced operating voltage [1], thus addressing the obstacle of power dissipation [4]. However, III-V films on Si are prone to formation of defects, including threading dislocations (TDs), because of lattice mismatch at the hetero-interface. Given the detrimental effects of such atomic defects on the electronic and photonic response of the III-V films, it is of crucial importance to develop techniques for detecting defects. This work presents a diagnosis to detect sub-micron optical signatures of sub-surface defects by collecting and selectively filtering, femtosecond-laser-generated second-harmonic generation (SHG) radiation from the sample through a 50 nm aperture fused silica probe. When the probe is scanned, samples laden with sub-surface defects uniquely exhibit prominent "hotspots" (i.e., locally intense SHG signal) due to near-surface intensity spikes created by scatter and localization of the fundamental field from the buried defects. The density, size, and pattern of these hotspots depend on the defect density and wavelength of the incident fundamental light but are not affected by and uncorrelated with the topology of the sample surface. Control samples without defects (*e.g.*, homo-epitaxial GaAs-GaAs films) lack such defect-related signatures. We observe no comparable signature using conventional linear optical probe microscopy, which is instead dominated by surface reflection and topology, nor using conventional far-field reflected SHG without probe collection, which is dominated by anti-phase-domains at the film surface. Thus, second-harmonic probe microscopy (SHPM) relies on both second-order nonlinearity and sub-wavelength collection can uniquely detect buried defects in semiconductors films noninvasively and requires no sample preparation. In contrast, most traditional techniques for diagnosing sub-surface defects require special, often invasive, sample preparation: e.g., electron microscopy and crystallographic etches [5,6]. Some such as Bragg coherent diffraction [7] require access to a large-scale light-source facility.

**RESULTS AND DISCUSSION:**
**Second Harmonic Probe Microscopy by Light Scattering and Localization**
Schematic of Fig.1A shows our approach for SHPM. Near-infrared pulsed laser penetrated the film and scattered from sub-surface defects. Semiconductor III-V films like GaAs on Si (001) with 4.2 % lattice mismatch at the interface, have a high density of TD defects which penetrate through the ~500nm top-layer GaAs and reach the surface of the sample. They appear as contrast [8] in cross-sectional scanning transmission electron microscopy (STEM) images e.g., Fig. 1B (bottom). Light localization is due to interference among multiply elastic backscattered waves [9-12] that creates a network of intensity spikes and nodes of sub-wavelength lateral dimension within the film. The interest in manipulating photons analog of electrons has resulted in some sub-wavelength influential discoveries such as Anderson light localization [9,13], nearfield optics [14], and optical transmission through sub-wavelength hole [15]. Light localization has been observed in non–conductive disordered media for photons [13, 16-18], conductive disordered media [19], nonlinear and disordered media [20,21], photonic crystal [22] and photonic moiré lattice [23]. Even localization in cold atoms in 3 dimensions and Bose-Einstein condensation [24,25] is analogous to light localization.

Since SHG intensity $$I_{2\omega} \sim [\chi^{(2)}]^2 I_\omega^2 \quad (1)$$

where $\chi^{(2)}$ is the local second-order nonlinear optical susceptibility, depends quadratically on fundamental wave intensity $I_\omega$, the SHG process enhances the contrast between local spikes and nodes at fundamental intensity. SHG filtering eliminates the strong laser reflection from the surface that could washed out the weaker backscattered and localized fundamental light. With a fiber-based probe scanning ~30 nm above the sample surface we collected back-propagating SHG light. These latter generated images show local SHG intensity variations with transverse resolution determined by aperture size and scan step size. This microscopy system includes a fiber probe, filters, and photo multiplier tube suppressing fundamental light by a factor of $10^{22}$ compared to SHG.

Fig. 1B (top) shows a SHG hotspot pattern from a 500 nm GaAs film grown on a Si substrate off-cut 4º from (001). This off-cut creates terraces separated by double-atomic-height steps that eliminated anti-phase domains (APDs). However, TDs density, determined from planar and cross STEM images, is $10^9$ cm$^{-2}$. A similar scan of a reference GaAs-GaAs sample free of any sub-surface defects, on the other hand, yields a uniform SHG signal free of high-contrast texture (Fig. 1c top). The maximum intensity of hotspots in the GaAs-Si-off axis scan exceeds SH intensity from the reference film approximately four-fold. This increase in intensity of reflection from disordered media has been previously attributed to Anderson localization of light [26].

Several cross-checks confirmed our interpretation of the SHPM signal including checking the quadratic dependency of SHG signal on fundamental excitation signal. We confirmed that the spectrum of the collected signal contained only frequency-doubled fundamental light (Supplementary Data Fig. 1a). This ruled out the possibility that fundamental light could leak through the fiber probe shaft and become the source for different frequency light. We checked a pre-fabricated sub-surface submicron-scale patterns e.g., alternating strips sample of InP and SiO2, and observed matching SHPM patterns (Supplementary Data Fig. 1B). Also, we checked the reliability of SHG features by scanning neighboring regions independently and showing optical hotspots matches on two sides (Supplementary Data Fig. 1C).

All the observed hotspot profile patterns have exponentially decay tails/wings, as expected for Anderson localization [13,20] (Supplementary Data Fig. 2A). The observed optical hotspot intensity distribution also shows a log-normal distribution which is the signature of localization [27,28] (Supplementary Data Fig. 2C).

**Subsurface Detection**
Sub-micron bright and dark spots such as those shown in Fig. 1B, could also arise from local defect-related variations in the film's $\chi^{(2)}$, rather than from variations in $I_\omega$. To test this possibility, we tuned the incident wavelength. A fixed spatial pattern of $\chi^{(2)}$ variations should retain a fixed shape as the excitation wavelength changes, even though the intensity of individual features may vary. On the other hand, the locations of hotspots in the backscattered pattern depend on interference, and thus on wavelength. Fig. 2 shows two examples of how the observed SHPM pattern evolves as the excitation (collection) wavelength tunes from 780 (390 nm) (A, D) to 840 (420 nm) (B, E). Fig. 2A and 2B are raster scans of the same 2x2 µm area of a GaAs (500nm)-Si (001)-on axis sample, Fig. 2D and 2E of a GaAs (500nm)-Si (001):4º vicinal sample. In both cases, the location of hotspots is completely different at these two wavelengths. At intermediate wavelengths (not shown here), we observe that the patterns evolve continuously. In either case, however, we do not observe any correlation between the hotspot patterns and surface topographies, shown in Figs. 2C and F, which correspond to rms roughness 2 and 16.5 nm, respectively, for the on- and off-axis samples.

By replacing the SHG bandpass filters on the detector with filters at the fundamental wavelength, we directly compare fundamental light scans with SHG scans. Figs. 2 G-I show, respectively, a fundamental light (780 nm) scan, surface topography (derived from the probe feedback signal), and SHG scan of the same 2x2 µm area of an oriented, 500 nm GaAs film grown on Si(001). The dominant features of the fundamental light scan (Fig. 2G) closely match the main surface topographical features (Fig. 2H). These linear features are also independent of wavelength: we observe them whether we illuminate the sample with fs-pulsed or continuous wave near IR light, incoherent white light, or frequency-doubled (390 nm) pulses (not shown). In contrast, the dominant features of the SHG scan (Fig. 2I) obtained simply by switching filters and collecting only 390 nm light, are uncorrelated with either surface topography (Fig. 2H), or with secondary features visible in the fundamental light scan (Fig. 2G). Like the SHG scans of GaAs films (Fig. 2A and 2D), the scan in Fig. 2I depended strongly on wavelength. We observe similar trends on a wide variety of samples. Evidently, strong fundamental reflection from surface features masks the weak hotspots from sub-surface defects, whereas SHPM better discriminates the latter.

To test the hypothesis that SHG hotspots originate from *sub*-surface structures, we prepared a series of GaAs/Si(001) or InP/Si(001) samples with groove-like submicron-scale aspect-ratio trapping (ART) structures at the buried interface. These structures consist of parallel 170nm high SiO2 pillars fabricated on the off-cut Si(001) substrate along the [110] direction, separated by ~ 90 nm trenches, which GaAs or InP was deposited. Fig. 2L shows a typical SHPM scan of such a sample. The grooved topology of the buried interface, though not precisely mapped, is clear evident in the SHG pattern, in contrast to those like Fig. 2I. Fundamental wavelength scans (Fig. 2J), on the other hand, are dominated by surface topography (Fig. 2K). This demonstrates that sub-surface structure is responsible for the broad features of the SPHM pattern. Inset SEM graph in Fig. 2K shows the strip signature of these ART structures.

**Electric Fields Simulation**
To better understand the structure of experimental SPHM patterns, we emulated near-field light scattering in III-V thin films using finite element calculations. These simulation data show GaAs-Si structures without dislocation scattering site (Fig. 3A) low density of dislocation scattering sites (Fig. 3B) and full of sub-surface scattering sites (Fig. 3C) that scatter light similarly to experimental results shown in Fig. 1 and 2. Dislocation defects behave as acceptor traps for electrons and thus act as Coulomb scattering centers [29]. To emulate their scattering behavior, we assign them locally metallic properties. This leads to scattering and localization of light electric fields by scattering sites and creating patterns of sub-micron hotspots (Fig. 3C) outside (and as well inside) the III-V film due to interference which resemble the observed SHPM plots. A lower density of scattering sites (Fig. 3B) yields broader hotspots than a higher density, as observed in experiments.

**Correlation between Hotspots Size and Dislocation Defects Density**
We have applied the SHPM to a wide variety of III-V epitaxial films, with surface defect densities $n_d$ (dislocation defect densities-estimated from STEM images at surface of the films) ranging from $10^7$ to $>10^{10}$ cm$^{-2}$. A global trend, illustrated by the data in Fig. 4, is that higher defect density yields more densely packed hotspots. The left columns show SHPM scans, and representative profile cuts, of GaAs-Ge-off axis (Fig.4 A), GaAs-Si-off axis (Fig. 4B) GaAs-Si-on axis (Fig.4C) and In$_{30}$Ga$_{70}$As-GaAs (Fig.4D). The right column shows corresponding cross STEM, in which TDs appear as white streaks. Main plots (Fig. 4E) shows average FWHM of SHG hotspots vs. $n_d$. Over the range $10^7 < n_d < 10^{10}$ cm$^{-2}$, hotspot size scales roughly logarithmically with $n_d$: a three-decade increase in $n_d$ reduces SHG spot size by a factor of three. The trend is robust over a wide range of material systems. The characteristic length (hotspots separation) of sample with dense dislocation density (Fig. 4C) is much shorter comparing to sample with lower defect density (Fig. 4A).

Mean free path for photons is the average distance travelled between collisions at scattering sites. Measuring this characteristic length helps us understand Ioffe-Regel localization condition. We have used the recorded size of SHG hotspot in 2D as localization length by fitting an exponential function to these optical spots and later extract the optical mean free path from this. Multiple scattering and interference of the optical and electronics waves by random disorder (here dislocation defects) altering the eigenstate from being extended to localized state [9,11,21,30,31]. Optical mean free path can be extracted from

$$R_{\text{localization}} = L_{\text{mfp}} e^{\pi kL/2} \quad (2)$$

Where R is wave localization length, L is mean free path and k is wavenumber [28]. Fig. 4F shows the theoretical extracted optical mean free path for all measured film with variety defects density vs optical localization length (size of the hotspots). The extracted mean free paths are very similar to the average distance between dislocation defects sites in the whole film estimated from cross STEM images of the samples. As an example, for GaAs-Ge on axis sample the extracted theoretical optical mean free path is 29 nm while the average distance between dislocations in the whole film is 37 nm. This distance only at the surface of the sample with much lower defect density is 626 nm. Fig. 4F shows clearly that mean free paths are much smaller than optical wavelength thus satisfying the Ioffe-Regel localization condition. The larger the density of dislocations the smaller localization length (optical hotspots size) thus smaller mean free path. Description of the localization length and mean free path estimation and calculation brought in Supplementary Data 3.

The In$_{30}$Ga$_{70}$As-GaAs samples (Fig. 4D) illustrate an extreme example of a film with such high $n_d$, that our system did not resolve any SHG features. Thus, $n_d \sim +10^{10}$ cm$^{-2}$ appears to be an upper limit. We learned from the simulation data that extreme dislocation density with a gap below ~20 nm between the defects (like this In$_{30}$Ga$_{70}$As-GaAs sample) would block light from entering the film. Simulation (Supplementary Data Fig. 4) shows the random dense dislocation defects block the light from penetrating into the film. In addition, we studied the size and intensity of the hotspot in a GaAs-Si film as function depth of the film while sputtering the film. Experimental data (Supplementary Data Fig. 5) shows that the size and intensity of hotspots decrease as we reach the area with higher defect density close to GaAs-Si interface. This confirms our observation and simulation results that high density of dislocations effectively prevent light from penetrating inside the film.

SHPM patterns were not only sensitive to defect density but also to defects orientation. To illustrate this, we compared SHPM patterns of InGaAs-InP–GaAs-Si structures grown with and without a Tellurium (Te) surfactant, which served to relax strain within the InGaAs layer during growth. STEM image (Supplementary Data Fig. 6A) shows that dislocations within InGaAs layers grown with surfactant were along surface normal and yielded streaked SHPM patterns. Less relaxed InGaAs films grown with no surfactant, on the other hand, featured crossed and tilted dislocations as it shown in STEM micrographs (Supplementary Data Fig. 6B) and create localized looking hotspots as they appear in SHPM scan.

**CNOCLUSION:**
The ability to detect scattering and localization signatures of crystallographic dislocation defects address more details about the nature of light localization and offer an alternative path to light localization and super-resolution imaging compared to plasmonics. By using a nonlinear optical approach and high resolution apparatus we were able to clearly detect signatures of atomic-scale dislocation defects. This includes a bench-top technique that screens heteroepitaxial films for sub-surface dislocations, and provides qualitative indicators of defect density, orientation, and arrangement. Since it requires neither sample preparation nor contact, this approach could help guide the choice of defect-control strategies (e.g., surfactants, growth rate, substrate temperature) *in-situ* and in real-time during thin-film crystal growth. Strategic choice of sub-band-gap wavelength, which ensures that the incident light penetrates the buried defect origin site, should expand the applicability of SHPM to a wide range of technologically important $\chi^{(2)}$ films, including all III-Vs, strained silicon, [32] and ferroelectrics [33]. Moreover, since the basic process relies on the localization of defect-scattered *fundamental* light, *third*-harmonic probe microscopy could potentially characterize defect-laden films with centrosymmetric crystal structure in the same way, thus expanding applicability of the technique beyond $\chi^{(2)}$ materials to virtually any type of semiconducting or insulating film.

# Main References:

1. Del Alamo, J.A. Nanometre-scale electronics with III–V compound semiconductors. *Nature* **479**, 317-323 (2011).

2. Awschalom, D. D., Bassett, L. C., Dzurak, A. S., Hu, E. L., Petta. J. R. Quantum spintronics: engineering and manipulating atom-like spins in semiconductors. *Science* **339**, 1174-1179 (2013).

3. Tomioka, K., Yoshimura, M., Fukui., T. A III–V nanowire channel on silicon for high-performance vertical transistors. *Nature* **488**, 189-192 (2012).

4. Theis, T. N. & Solomon, P. M. It's time to reinvent the transistor!. *Science* **327**, 1600-1601 (2010).

5. Bollmann., W. Interference effects in the electron microscopy of thin crystal foils. *Phys. Rev.* **103**, 1588-1589 (1956).

6. Richards J. L. & Crocker. A. J. Etch pits in gallium arsenide. *J. Appl. Phys.* **31**, 611-612 (1960).

7. Clark, J. N. et al. Three-dimensional imaging of dislocation propagation during crystal growth and dissolution. *Nat. Mater.* **14**, 780-784 (2015).

8. Williams, D. B. & Carter, C. B. Transmission electron microscope. A textbook for materials science (Springer, Boston, MA, 2009).

9. John, S. Localization of light. *Phys. Today* **44**, 32-40 (1991).

10. van Tiggelen, B. A. Multiple scattering and localization of light. (1992).

11. Lagendijk, A., van Tiggelen, B. A., Wiersma, D.S. Fifty years of Anderson localization. *Phys. Today* 62, 24-29 (2009).

12. Wiersma, D. S. Disordered photonics. *Nat. Photonics* **7**, 188-196 (2013).

13. Wiersma, D. S., Bartolini, P., Lagendijk, A., Righini, R. Localization of light in a disordered medium. *Nature* **390**, 671-673 (1997).

14. Betzig, E. & Trautman, J. K. Near-field optics: microscopy, spectroscopy, and surface modification beyond the diffraction limit. *Science* **257**, 189-195 (1992).

15. Ebbesen, T. W., Lezec, H. J., Ghaemi, H.F., Thio, T., Wolff, P. A. Extraordinary optical transmission through sub-wavelength hole arrays. *Nature* **391**, 667-669 (1998).

16. Mascheck, M. et al. Observing the localization of light in space and time by ultrafast second-harmonic microscopy. *Nat. Photonics* **6**, 293-298 (2012).

17. Sperling, T., Buehrer, W., Aegerter, C. M., Maret, G. Direct determination of the transition to localization of light in three dimensions. *Nat. Photonics* **7**, 48-52 (2013).

18. Sapienza, L. et al. Cavity quantum electrodynamics with Anderson-localized modes. *Science* **327**, 1352-1355 (2010).

19. Smolyaninov, I.I., Zayats, A. V., Davis, C. C. Near-field second harmonic generation from a rough metal surface. *Phys. Rev. B* **56**, 9290-9293 (1997).

20. Segev, M., Silberberg, Y., Christodoulides, D. N. Anderson localization of light. *Nat. Photonics* **7**, 197-204 (2013).

21. Schwartz, T., Bartal, G., Fishman, S. and Segev, M. Transport and Anderson localization in disordered two-dimensional photonic lattices. *Nature* **446**, 52-55 (2007)

22. Soukoulis, C. M. *PHOTONIC CRYSTALS AND LIGHT LOCALIZATION IN 21ST CENTURY*. Springer Science & Business Media (2012).

23. Wang, P., Zheng, Y., Chen, X., Huang, C., Kartashov, Y.V., Torner, L., Konotop, V.V. and Ye, F., Localization and delocalization of light in photonic moiré lattices. *Nature* **577**, 42-46 (2020)

24. Aspect, A., Inguscio, M., Müller, C. A., Delande, D. Anderson localization of ultra cold atoms. *Phys. Today* **62**, 30-35 (2009).

25. Roati, G. et al. Anderson localization of a non-interacting Bose–Einstein condensate. *Nature* **453**, 895-898 (2008).

26. Sheinfux, H.H. et al. Observation of Anderson localization in disordered nanophotonic structures. *Science* **356**, 953-956 (2017).

27. Altshuler, B. L., Kravtsov, V. E., Lerner, I. V. Applicability of scaling description to the distribution of mesoscopic fluctuations. *Phys. Lett. A* **134**, 488-492 (1989).

28. Richardella, A. et al. Visualizing critical correlations near the metal-insulator transition in $Ga_{1-x}Mn_xAs$. *Science* **327**, 665-669 (2010).

29. Weimann, N. G., Eastman, L. F., Doppalapudi, D., Ng, H. M., Moustakas, T. D. Scattering of electrons at threading dislocations in GaN. *J. Appl. Phys.* **83**, 3656-3659 (1998).

30. Lee, P.A. and Ramakrishnan, T.V., 1985. Disordered electronic systems. *Rev. Mod. Phys.* **57**, 287 (1985)

31. Lee, M., Lee, J., Kim, S., Callard, S., Seassal, C. and Jeon, H. Anderson localizations and photonic band-tail states observed in compositionally disordered platform. *Sci. Adv.* **4**, e160279 (2018).

32. Cazzanelli, M. et al. Second-harmonic generation in silicon waveguides strained by silicon nitride. *Nat. Materials* **11**, 148-154 (2012).

33. Choi, K. J. et al. Enhancement of ferroelectricity in strained $BaTiO_3$ thin films. *Science* **306**, 1005-1009 (2004).

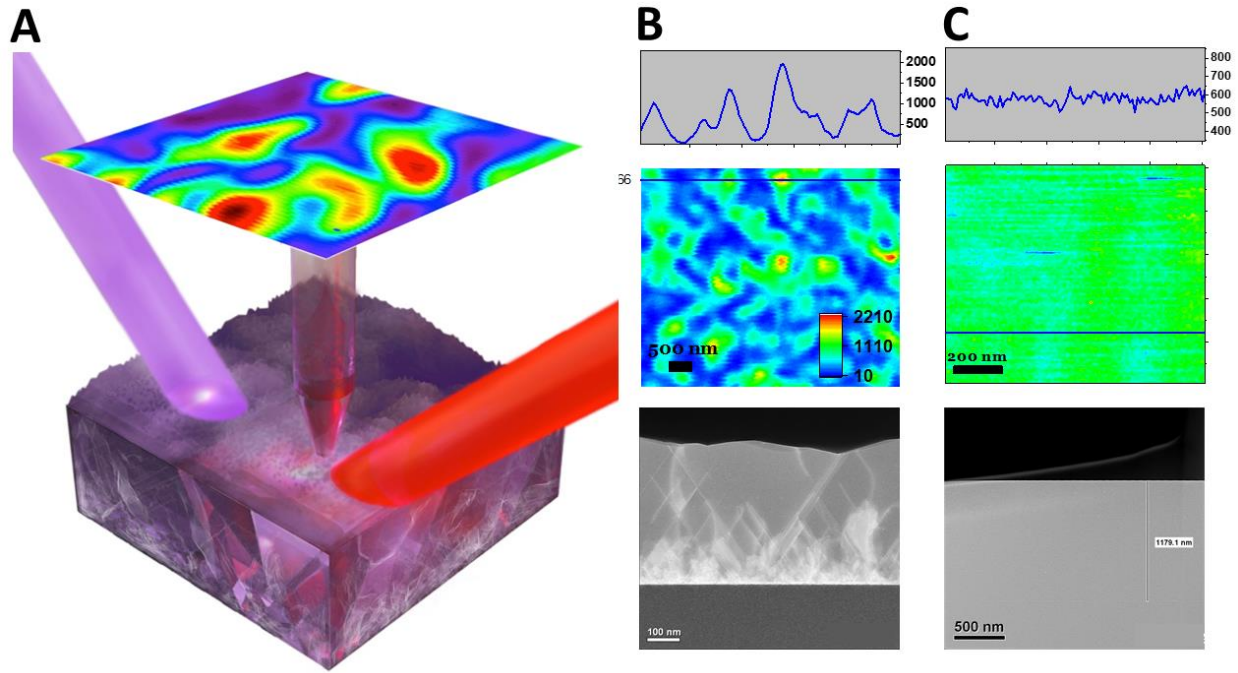

**Fig. 1 Nonlinear defect detection technique. (A)** Schematic of the setup shows how the probe microscope (SHPM) pick up the propagating light wave from defects area. Fundamental light reaches the defects area and get scattered and localized by multiple elastic backscatters. The SHG signature of localized light get collected by 50 nm uncoated aperture probe and filtered to block the intense fundamental light. The 2x2 μm topography and SHG signal collected at the same time from GaAs-Si-off axis show no correlation to each other. **(B)** GaAs-Si-off axis raster scan show clear SHG hot spots as a signature of the presence of defects in the film. The profile cut shows that the hotspots intensity are several times larger than the intensity of the bulk GaAs-GaAs sample. Cross STEM shows the penetration of defects from mismatch GaAs-Si interface to the film surface. The thin film top layer is 500 nm. **(C)** GaAs-GaAs sample raster scan and profile cut shows uniform intensity. Cross STEM shows there is no signature of any type of defects at GaAs thin film interface and top layer.

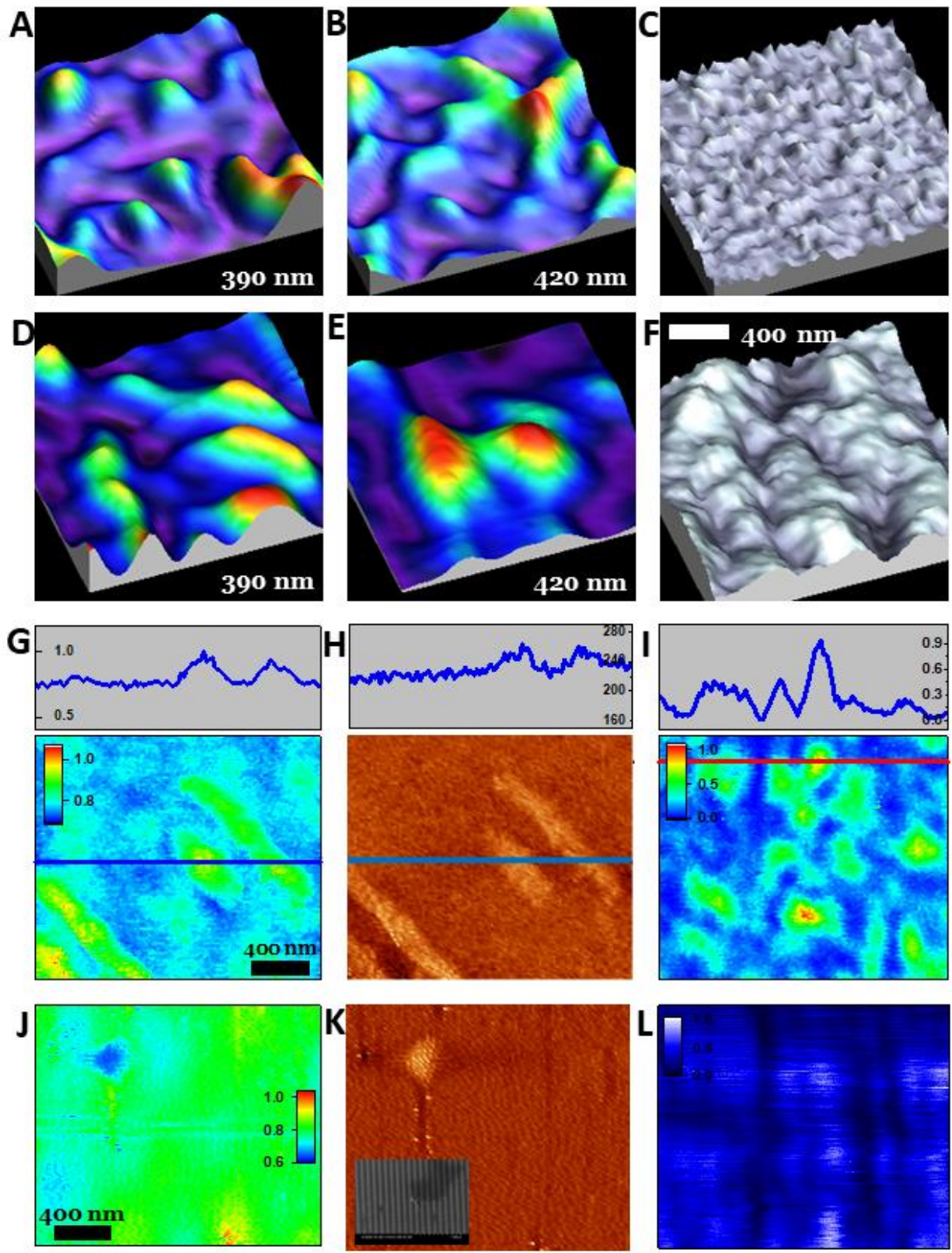

**Fig. 2 Evolution of SHG optical domains at different excitation/collection wavelength and Comparison of fundamental and nonlinear (SHG) optical scan. (A, B)** SHPM plot of GaAs-Si-on axis at 420 nm (laser excitation at 840 nm) (A) and at 390 nm (laser excitation at 780 nm) (B) collected by SHPM probe. **(D, E)** SHG optical plot of GaAs-Si-off axis at 420 nm (laser excitation at 840 nm) (D) and at 390 nm (laser excitation at 780 nm) (E). Evolution of optical domains is observed for both sample as a function of wavelength. All the steps of evolution are not shown here. These evolutions show that the variation of intensity is not due to variation of local $\chi^{(2)}$ and are strong signature of size dependency of the hotspots scattering sites. **(C, F)** Topography collected by scanning probe microscope during the optical scan are shown in gray color. There is no correlation between optical map and topography of the sample. GaAs-Si-on axis (C) has Rrms=~2 nm roughness. GaAs-Si-off axis (F) has Rrms= ~16.5 nm roughness. **(G-I)** Fundamental scan of GaAs-Si-on axis (G) shows strong correlation to the topography (H) which was collected at the same time by the feedback loop system that controls the sample-probe distance. Profiles cuts in (G) and (H) show how topography is dictating in linear light study. Profile cut for topography is in nm unit and for linear scan is in counts per 0.2 second. Excitation and collection were at 780 nm for linear study. Nonlinear scan (I) of the same area show signature of localization of the light at defects area beneath the surface of the film. Excitation and collection were at 780/390 nm for SHG study. **(J-L)** To check the concept of subsurface defects contribution to this SHG signal, we have prepared and used a series of sample with subsurface SiO2 and InP or GaAs ART pattern. The linear optical study(J) did not show any signature of these subsurface structures while dominated by topography (K). SHPM optical study (L) resolve the strip pattern. The inset picture in topography plot (K) is SEM image of ART sample. SHG intensities are normalized (G-L) for comparison.

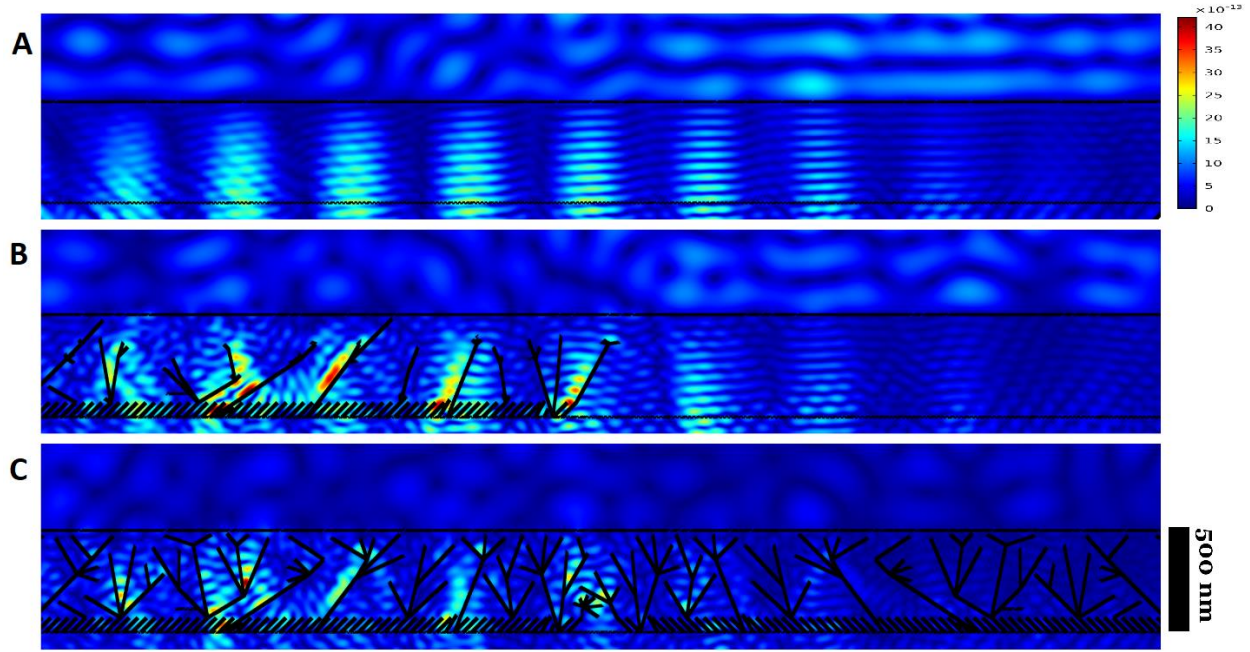

**Fig. 3 Finite elements analysis over light (electric fields) scattering by dislocation scattering sites in Air-GaAs-Si media.** Fundamental linear excitation light at 780nm get scattered and localized by dislocation scattering sites inside the GaAs film. SHG signal due to nonlinearity of GaAs was calculated for the film without dislocation **(A)**, with low density of dislocation **(B)** and high density of dislocation **(C)**. Introducing random dislocation defect scattering sites, would introduce hotspot looking scatted light(electric fields) just outside the film similar to experimental observations in Fig. 1 and 2. Lower density scattering sites (B) introduce broader hotspots comparing to higher density film (C) while the electric fields looks more uniform in empty area of the film (B). SHG scattering and localized light in GaAs medium without any dislocation scattering sites (A) shows almost uniform intensity as it was observed in experiment. The GaAs film was excited by 780 nm light with 45° respect to surface normal. Excitation spot was 2500nm area on left side of the film only.

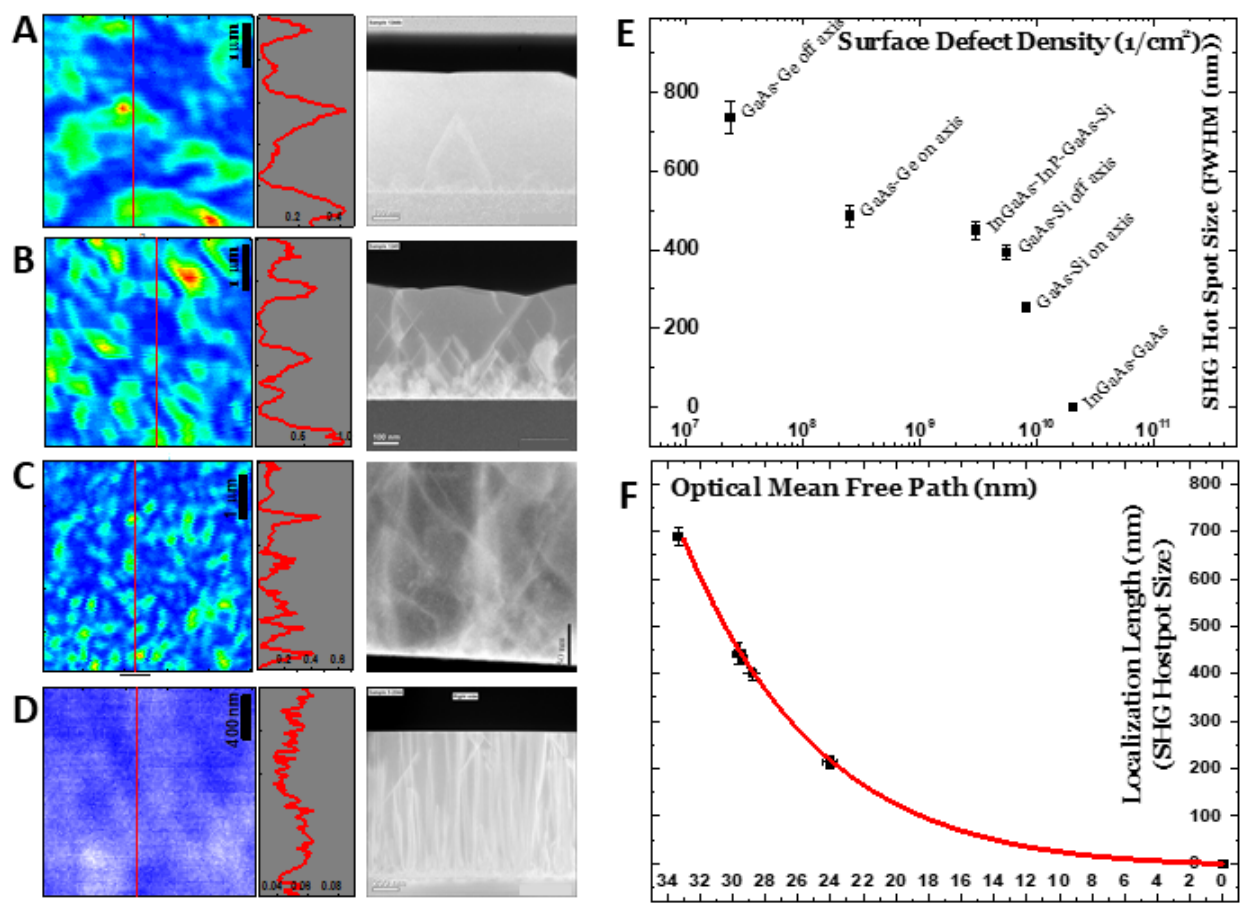

**Fig.4 Optical hotspot size correlation to defect density and optical mean free path.**
The size (FWHM) of hotpots were extracted from SHG plot left columns. The left columns show the SHG optical scan collected by SHPM system over GaAs-Ge-off axis **(A)**, GaAs-Si-off axis **(B)** GaAs-Si-on axis **(C)** and $In_{30}Ga_{70}As$-GaAs **(D)** and the profile cut used for hot spot size (FWHM) measurement. The right columns show cross STEM from each thin film clearly show the TD defects for each film. These STEM images show low defect density in GaAs-Ge film (A), moderate defect density in GaAs-Si-off axis (B), high defect density in GaAs-Si-on axis (C), comparing to extremely high defect density in $In_{30}Ga_{70}As$-GaAs (D). **(E)** This Quantitative SHG hotspot size study in variety of thin films with different surface defect density (extracted from STEM) had shown a decreasing trend in size of optical hotspots while TD defect density was increasing. **(F)** The size of 2D optical hotspot as localization length was used based on Eq. (2) to calculate optical mean free paths. Optical mean free path as average distance for travelling photons while scattered by dislocation defects is related to average distance between the defects. The mean free paths are fraction of excitation wavelength showing satisfaction of Ioffe-Regel localization condition. The fitting red line is theoretical relation between these two parameters by Eq. (2).

# Methods:

## a) Experimental Method and Setup

SHG optical study of semiconductor thin film was performed by fiber based nonlinear nearfield scanning optical microscope (NSOM) system with 50 nm fiber aperture (Fig. 1a schematic). This uncoated probe scanning approach was used to avoid and minimize any enhancement and perturbation of the electromagnetic field at probe area [19]. The probe was kept at ~ 30 nm above the sample with feedback loop system monitoring the amplitude of the scanning probe. 76 MHz laser with ~150fs pulse width at ~780nm was focused on area about 5-10 µm on the sample (not in scale in Fig.1) and collection was done at 390nm. The incident angle is ~45°. The Sample got scanned by piezoelectric stage under the stationary probe. Fiber probe which support light only below 600nm pick up the propagating SHG signal of III-V film at nearfield regime and signal get filtered for residue of fundamental light. Photoelastic modulator tube (PMT) sensitive only to photon at 200-700 nm range was used with photon counter system to measure the photons. The nonlinear response of the GaAs-Si-on axis film is typically ~$10^{15}$ times weaker than linear reflection of the film. Nonlinear scanning probe microscope system has advantages of having very high spatial resolution and being noise free by collecting the SHG signal away from the linear signal.

Study in the linear regime with excitation at 780nm has a dis-advantage of not being able to distinguish the very weak scattered and localized light which create those hotspots. If we use the probe microscopy to look at the same wavelength of excitation light, the reflection of the excitation light at the surface dominate all the intensity information and the weak scattered and localized light at sub-surface dislocation area would not be distinguished. Instead of the linear study, if we look at the nonlinear response of the III-V film at 390nm, then there would not be such a problem of dominating reflection light from surface of the sample as the only reflected and dominating signal at surface is 780nm which is filtered. Then scattered and localized SHG light intensity can be distinguished from the film typical SHG background respond. This filtering approach is capable of distinguishing the very weak SHG scattered and localized hotspot at presence of dominating surface reflection.

## b) Sample Growth

The III-V film such as GaAs was grown on on-axis Si (001) by metal-organic chemical vapor deposition (MOCVD) technique using two-step growth approach. AIXTRON CRIUS-R MOCVD system was used for that purpose. Essential silicon wafer cleaning and hydrogen passivation was done by vapor HF and wet HF processes. To promote the formation of double steps on Si along <110> direction for prevention of antiphase domains, baking at high temperature (>800 C) was performed. III–V films were grown by using trimethylindium (TMIn) and trimethylgallium (TMGa) as the group-III precursors, tertiarybutylarsine (TBAs), tertiarybutylphosphine (TBP) as well as arsine (AsH3) and phosphine (PH3) as the group V precursors. To have charge neutrality along the interface and promote the growth of single domain GaAs, wafer surface was saturated with an arsenic monolayer by introducing TBAs in the reactor at low temperature (<500 C). Two step growth was introduced by <20 nm GaAs LT nucleation layer at (<450 C) by low V/III ration with roughness ~1nm measured by AFM and SEM. A 500nm thick GaAs was grown at ~600 C by using AsH3 with high V/III ratio and growth rate of ~1.3 micrometer/h with ~0.6nm roughness.  Quality and defect density of the crystal was checked by a high-resolution X-ray diffraction (HRXRD) and cross and planar STEM later. Annealing had performed at the end to improve the quality of the crystal ay 750C [34].

**Method and Supplementary Data References:**

34. Orzali, T. et al. Growth and characterization of an In 0.53 Ga 0.47 As-based Metal-Oxide-Semiconductor Capacitor (MOSCAP) structure on 300mm on-axis Si (001) wafers by MOCVD. *J. Cryst. Growth* **427**, 72-29 (2015).

35. Lerner, I. V. Distribution functions of current density and local density of states in disordered quantum conductors. *Phys. Lett. A* **133**, 253-259. (1988).

36. Goodman, G. W. Some fundamental properties of speckle. *J. Opt. Soc. Am.* **66**, 1145-1150 (1976).

37. Dainty, J. C. & Christopher, J. *LASER SPECKLE AND RELATED PHENOMENA*. Springer Science & Business Media (2013).

**Acknowledgments:**
The work was supported in part by Welch Foundation grant F-1038 and The National Science Foundation. The authors acknowledge the helpful discussions with the following: Matthew Foster and Brian DeMarco over light localization; Tom Schamp over STEM images; Agham Posadas over thin film diagnosis; Ricardo S. Decca over scanning probe optical microscopy; Andrei Dolocan over secondary mass ion spectroscopy sputtering data, and Raluca Gearba, Karalee Jarvis, Hugo Celio and Richard Piner at CNM/TMI centers over electron microscopy data. We thank Xiaoqin Li's group for accessing to AFM for the preliminary study.

**Authors Contributions:**
F.S. designed and developed the setups, performed the experiments, analysed the data and conceived the idea. T.O., A.V., P.H. and M.W. prepared and characterised the samples. F.S. and M.M. performed the theoretical simulations. Results and analysis were discussed between all members and paper was written by F.S., A.A, G.B. and M.D. Theoretical works, samples growth and the experiments were supervised by A.A., G.B. and M.D.

**Competing Interests:**
The University of Texas at Austin has a pending patent under the name of inventors F.S. and M.D. (patent application US16157765) based on the technique described and performed in this experiment for detection of crystallographic defects.

**Correspondence and requests for materials:**
All requests should be addressed to F.S.

**Data availability**:
The data that support the findings of this study are available from the corresponding author upon reasonable request.

# Supplementary data for:
# Atomic-Scale Defect Detection by Nonlinear Light Scattering and Localization

**Authors:**
Farbod Shafiei [1,*], Tommaso Orzali [2], Alexey Vert [2,3], Mohammad-Ali Miri [4,5], P. Y. Hung [2], Man Hoi Wong [2], Andrea Alù [4,6,7], Gennadi Bersuker [8], Michael C. Downer [1]

**Affiliations:**
[1] Department of Physics, The University of Texas at Austin, Austin TX USA
[2] SEMATECH, Albany NY USA
[3] Sunny Polytechnic Institute, Albany NY USA
[4] Department of Electrical and Computer Engineering, The University of Texas at Austin, Austin TX USA
[5] Department of Physics, Queens College of the CUNY, Queens NY USA
[6] Photonics Initiative, Advanced Science Research Center, CUNY, New York NY USA
[7] Physics Program Graduate Center, CUNY, New York NY USA
[8] The Aerospace Corporation, Los Angeles CA USA

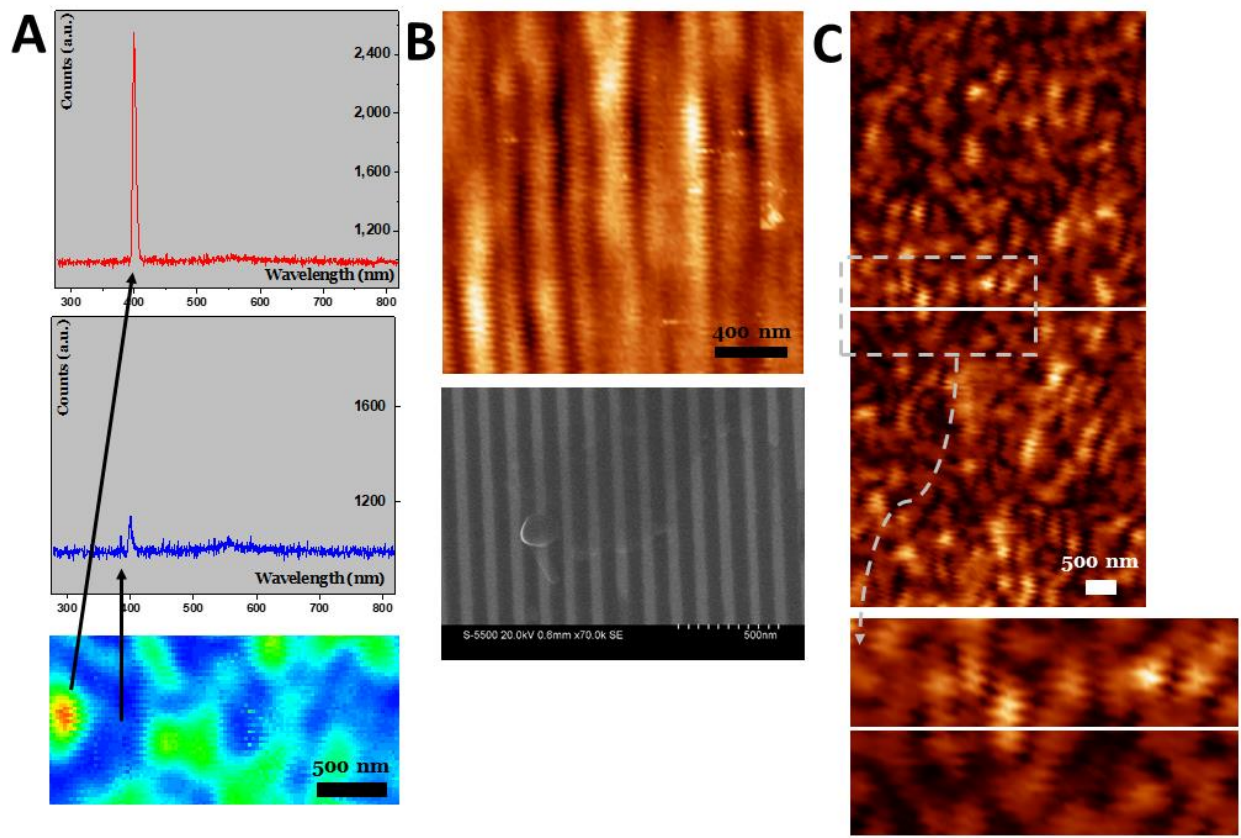

**Supplementary Data Fig. 1  Reliability of measured SHG signal.**
(**A**) Spectra of collected signals by SHPM probe from bright and dark areas of the GaAs-Si scanned SHG plots were studied. These spectra showed that there are no extra signals other than SHG (390 nm) signal that might leak into the fiber probe. Only excitation 790nm light was suppressed by a notch filter in this process. (**B**) SHG scan of a stripe pattern of ART structure (strip pattern of InP and SiO2 trenches with ~90 nm width) and its SEM image. The SHPM plots distinguish these strip patterns showing the reliability of technique in distinguishing a known pattern. (**C**) Scanned the neighboring regions of GaAs-Si film with moderate density of dislocations to compare the hotspots at boundary. The optical hotspots on both sides of the scanning area matched well showing the reliability of the measurement.

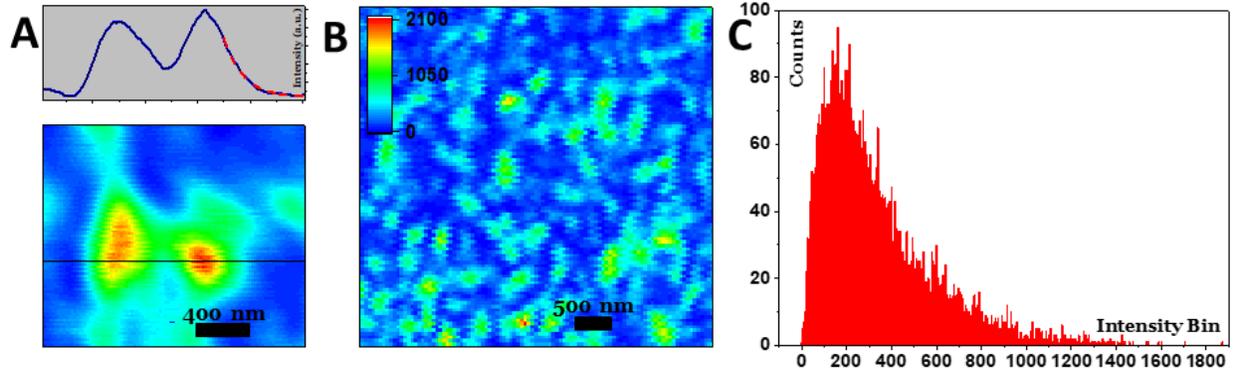

**Supplementary Data Fig. 2 Anderson localization signatures of SHG hotspot .**
Signatures of light localization of observed SHG hotspots have been checked which include shape of tail/wing of optical profile cut of hotspots and histogram of intensity distribution of hotspots. **(A)** The tail/wing of the profile cut of the observed hotspot belonging to GaAs-Si-off axis shows exponential decay tail (fitting in red stripe). Based on previous studies [13,20] ,Anderson localization has exponential decay tail/wing profile **(B)** Intensity distribution of GaAs-Si-on axis was studied. **(C)** The intensity histogram of hotspots shows a log-normal distribution. Based on original theoretical work [27,35], Anderson localization event has log-normal intensity distribution $\varphi(I) \propto \frac{1}{I} e^{-ln^2(I)}$. This intensity distribution is different from laser speckle intensity statistics with negative exponential distribution [36,37]. In addition, bulk samples without dislocation and hotspots show gaussian intensity distribution (not shown here).

**Supplementary Data 3  Estimating optical localization length and optical mean free path for different dislocation defect density medium.**

As Anderson localization was a "wave" interference phenomenon, later it has been extended to its counterpart electromagnetic waves and optics [9,20]. In our experiment dislocations acting as scattering sites brings the light to localization state. We used our 2D SHG hotspots size as an estimation for optical localization length by fitting an exponential function $I \sim e^{-2d/R}$ to these intensity plots. R is the localization length and d is the dimension variable. We had estimated these localization lengths for films with different density of dislocation defects. We approximated our calculation to 2 dimensions as our optical collection has 2D characteristics. This would let us to use the perturbative estimation over wave localization length function of realistic physical parameters including mean free path [30] $R_{localization} = L_{mfp} e^{\pi k L/2}$ as expressed in Eq. (2) in the paper. The scattering and localization had happened in fundamental light and we had collected the light localization as hotspots in the SHG regime. Using this nonlinear SHG optical dimensions instead of fundamental light dimensions is a close approximation as we do see in our simulation data that the fundamental hotspots are 2-3 times larger than SHG hotspots, balancing the difference of wavelength used in estimation. After estimating the SHG hotspots dimension and calculating the mean free paths by Eq. (2), we observed that optical mean free paths are a portion of wavelength satisfying Ioffe-Regel condition as shown in Fig. 4f. Physical evidence of the optical mean free path was estimated by calculating the average distance between dislocation defects in the whole sample from cross STEM images. The average distances between dislocations are very similar to this calculated optical mean free path for a variety of samples.

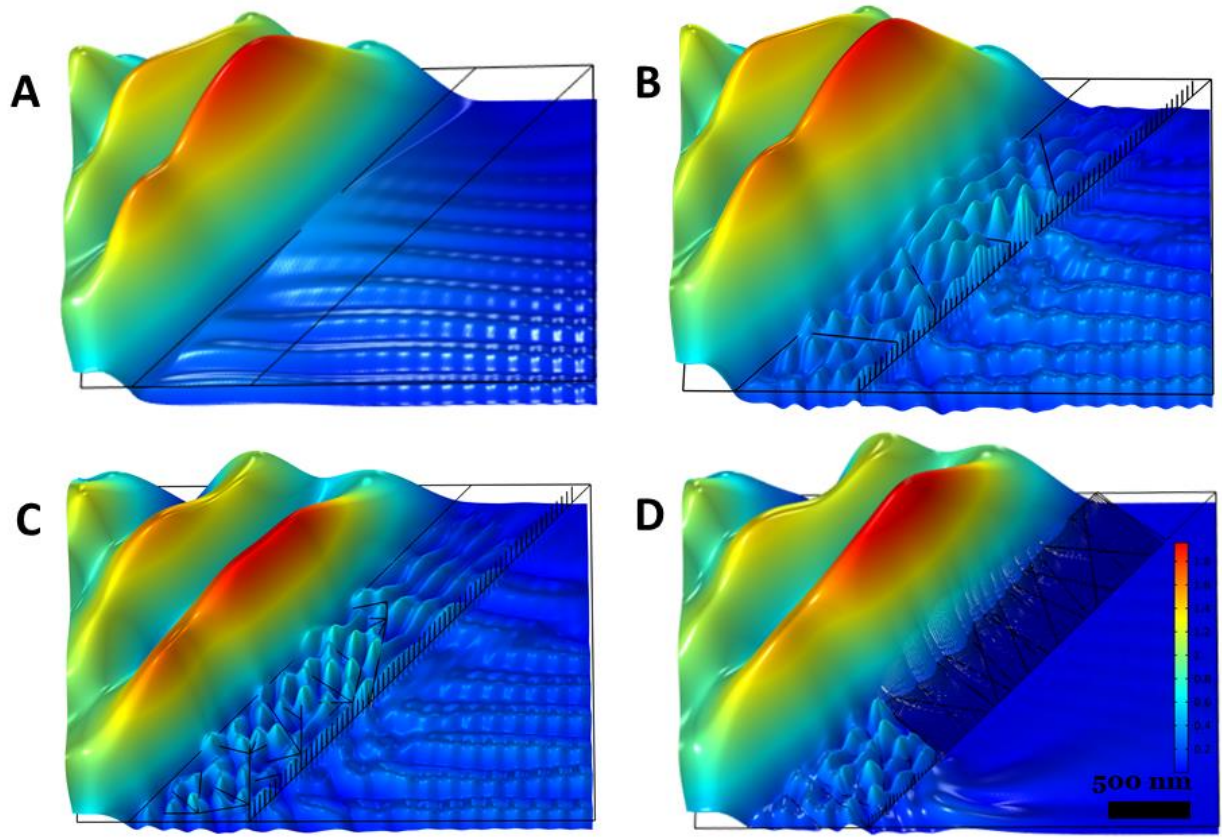

**Supplementary Data Fig. 4 Blocking the light by high density of dislocations.**
Simulation study shows that the fundamental excitation light gets blocked from penetrating into the GaAs-Si film when the density of dislocation close to the surface of the sample surpasses a high density threshold in a random order. The higher density of dislocation produces narrower fundamental hotspot (resembling SHG hotspot features observed in experimental result in Fig. 1 and 2 in main text. **(A)** GaAs film without any scattering sites (dislocations) shows a semi-uniform intensity inside the film. The GaAs sample makes 45 degrees with excitation linear 780nm light**. (B,C)** Lower/higher random density of dislocation defect scattering sites, would introduce broader and narrower hotspots inside the GaAs film. **(D)** Light gets blocked out from penetrating to the film as the light backscatter at the surface of the film where defects gap between neighboring dislocations become less than 20 nm in random order. This was observed in experimental SHG study of the films with very high density of dislocations in Fig. 4D in the main text.

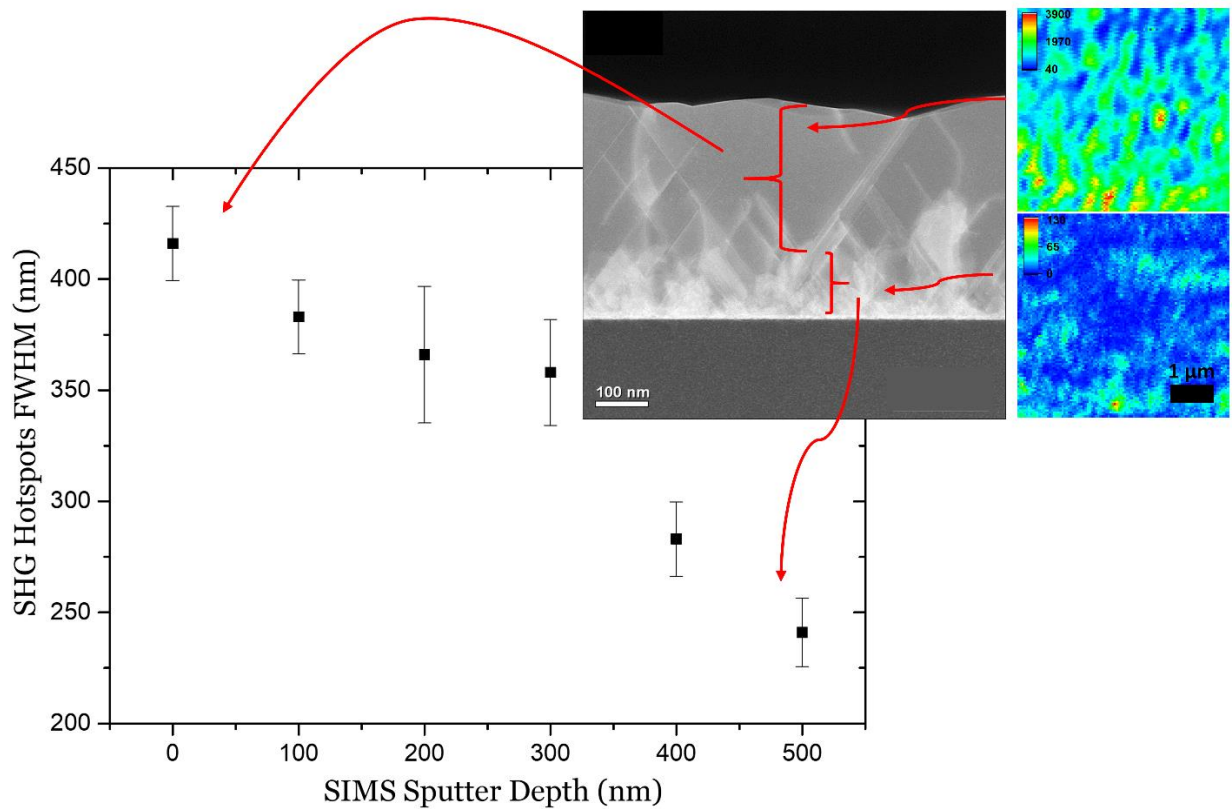

**Supplementary Data Fig. 5 SHG hotspot size and intensity study as function sample depth.**
The optical nonlinear measurement (average over all pixel of a 5x5µm scanned by SHPM of a GaAs-Si film with 500nm thickness) show a decrease in size and the intensity of hotspots areas as we get closer to GaAs and Si interface where the film defect density is very high. The cross STEM shows that in this GaAs-Si interface region there is no room for the light to scatter and localize while in the area close to the surface of film there is enough room for light localization. Secondary ion mass microscopy (SIMS) was used to sputter the sample with 100nm steps and then SHPM was used to scan each depth. Top SHG plot shows the presence of hotspot at the surface of the film while bottom SHG plot shows there is very little hotspot at the area very close to GaAs-Si interface with extremely low intensity. This shows that the area with dense defect density could decrease the penetration of the light.

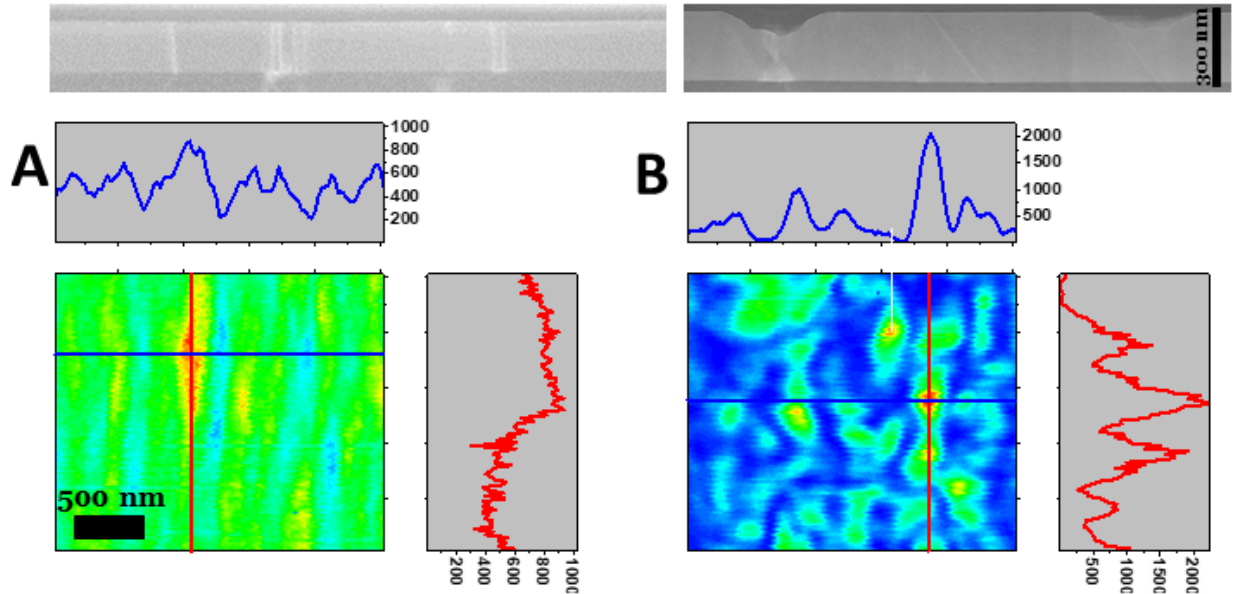

**Supplementary Data Fig. 6 | Sensitivity of the technique to the orientation and arrangement of defects growth.**
InGaAs-InP-GaAs-Si film with different dislocation orientation had been studied by nonlinear probe microscopy. **(A)** The InGaAs-InP–GaAs-Si sample has been grown under the controlled temperature and rate with Te dopant (as surfactant) to have a more relaxed crystal. In the upper panel, the STEM image of InGaAs top layer on top panel shows that in this film the orientations of dislocations are almost perpendicular to the InGaAs-InP interface. The SHG optical plot in the lower panel shows a streak looking pattern for this film. **(B)** The InGaAs-InP-GaAs-Si sample with different temperature and rate and without any surfactant dopant has been grown in less relaxed condition. The STEM image in the upper panel shows the higher defect density and some pits which is a clear signature of not a relaxed growth. The x-crossing orientation of dislocation defects is clearly different from the relaxed film. The tilted and random orientation growth of the defects created more cavity looking spots. The SHG optical plot of this film shows clearly more localized and more intense hotspot. This shows that SHPM technique is not only sensitive to density of scattering dislocation defects but also to their growth orientations